\documentclass[aps,twocolumn,amsmath,amssymb,floatfix,showpacs,
superscriptaddress]{revtex4}

\usepackage[dvips]{graphics}
\usepackage{color}
\definecolor{gold}{rgb}{0.85,0.66,0}
\definecolor{dred}{rgb}{0.6,0,0}

\begin{document}

\title{\textcolor{dred}{Determination of Rashba and Dresselhaus 
spin-orbit fields}}

\author{Santanu K. Maiti}

\email{santanu.maiti@saha.ac.in}

\affiliation{Theoretical Condensed Matter Physics Division, Saha 
Institute of Nuclear Physics, Sector-I, Block-AF, Bidhannagar, 
Kolkata-700 064, India} 

\affiliation{Department of Physics, Narasinha Dutt College, 129 
Belilious Road, Howrah-711 101, India}

\begin{abstract}
Determination of Rashba and Dresselhaus spin-orbit interaction strengths 
in a particular sample remains a challenge even today. In this article we 
investigate the possibilities of measuring the absolute values of these 
interaction strengths by calculating persistent charge and spin currents 
in a mesoscopic ring. Our numerical results can be verified experimentally.
\end{abstract}

\pacs{73.23.-b, 73.23.Ra.}

\maketitle

\section{Introduction}

Determination of spin-orbit (SO) coupling strengths in meso-scale and 
nano-scale semiconductor structures has drawn a lot of attention since
it is extremely crucial for designing spintronic devices. SO interactions
couple the orbital motions of electrons to their spin and lift the
degeneracy between spin up and down states. It gives the possibility of
manipulating and controlling the spin of an electron rather than its 
charge~\cite{zutic,datta,ando,ding,bellucci}. In solid-state materials 
one comes across two types of SO fields, 
\begin{figure}[ht]
{\centering \resizebox*{6.5cm}{2.5cm}{\includegraphics{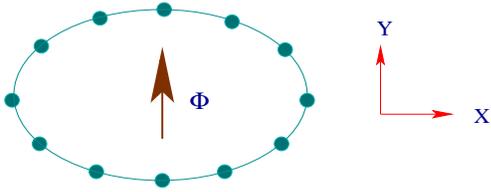}}\par}
\caption{(Color online). Schematic view of a mesoscopic ring threaded
by an AB flux $\phi$. The filled red circles correspond to the positions
of the atomic sites.}
\label{ring}
\end{figure}
namely, Rashba and Dresselhaus fields. SO interaction is a relativistic
effect. In the reference frame of a moving electron, an electric field is
converted to a magnetic field which induces a spin splitting through the
coupling of the electron spin with its momentum. Depending on the origin
of the electric field we refer SO field as a Rashba or Dresselhaus 
field. If the electric field is originated from a structural inversion 
asymmetry we get Rashba field. On the other hand, Dresselhaus field is 
obtained when the electric field is developed from a bulk inversion 
asymmetry. The precise measurement of both these two fields are 
extremely important to design spin based electronic devices.

Usually the strength of Rashba SO coupling is much higher than the 
Dresselhaus SO coupling, and therefore, people have initially tried to 
measure the Rashba term. In an experiment Koga {\em et al.}~\cite{koga}
have measured the values of Rashba coupling in quantum wells using the
weak antilocalization (WAL) analysis in terms of structural inversion 
asymmetry of the quantum wells. This WAL approach provides a useful
tool for determining Rashba SO coupling. Using photocurrent 
measurements~\cite{gani} on quantum wells, on the other hand, the 
relative strengths of Rashba and Dresselhaus terms can be very nicely 
deduced where angular distribution of the spin-galvanic effect at 
certain directions of spin orientation in the plane of a quantum well 
is used. A gate-controlled crossover from weak localization to 
antilocalization in coherent transport using a two-dimensional
electron gas has allowed to estimate separately the Rashba and 
Dresselhaus SO coupling terms~\cite{miller}. In a particular sample,
Rashba strength can also be monitored by applying electric fields
from gates~\cite{grun,jun} or by controlling the density of 
electrons~\cite{mat,hei} which generates a tremendous interest in the 
field of spintronics. Very recently, in a nice experiment Meier 
{\em et al.}~\cite{meier} have shown that both Rashba and Dresselhaus
SO fields can be measured by optically tuning the angular dependence 
of the electrons' spin precession on their direction of movement with 
respect to the crystal lattice.

Though some experiments are available for measuring Rashba and Dresselhaus
SO fields, yet the determination of these fields in a single sample, 
particularly within a tight-binding (TB) formalism, is still laking, to 
the best our knowledge. The present work tries to answer this issue. Here 
we propose the possibilities of measuring the absolute values of Rashba 
and Dresselhaus SO interaction strengths by calculating persistent charge 
and spin currents in a mesoscopic ring. A mesoscopic ring formed 
at the interface of two semiconducting materials is an ideal candidate 
for observing the interplay between two types of SO interactions. Due to
the band offset at the interface of two semiconducting materials an 
electric field is established which can be described by a potential 
gradient normal to the interface~\cite{premper}. This potential becomes
asymmetric, leading to the presence of Rashba SO interaction. On the
other hand, at such interfaces bulk inversion symmetry is naturally 
broken which gives rise to Dresselhaus term.

It is well known that a mesoscopic ring threaded by a magnetic flux, 
the so-called Aharonov-Bohm (AB) flux, carries persistent charge current.
B\"{u}ttiker {\em et al.}~\cite{butt} first studied the phenomenon of
persistent charge current in a mesoscopic ring and then several works
have been made to understand its behavior in single-channel rings and 
multi-channel cylinders~\cite{cheu1,alts,schm,ambe,ore,peeters,san1}. Later 
many experiments~\cite{levy,chand,jari,deb} have also been performed to 
justify the existence of non-decaying charge current in such systems.
Similar to persistent charge current in a mesoscopic ring induced by
an AB flux, the spin of an electron acquires a spin Berry phase while
traversing through a ring in presence of SO interaction results a 
persistent spin current~\cite{qfsun1}. A mesoscopic ring with SO coupling 
can provide persistent spin current even in the absence of any external 
magnetic flux or magnetic field. This is the so-called pure persistent 
spin current since it is not accompanying any charge current.

To determine the absolute values of Rashba and Dresselhaus SO fields
here we propose two different possibilities. First, by measuring 
persistent charge current in a mesoscopic ring in presence of magnetic
impurity. Second, by calculating persistent spin current in an
ordered mesoscopic ring. With these two approaches we can clearly 
estimate the values of SO coupling strengths.

In what follows, we present the results. In Section II, we describe
the model and theoretical formulation to calculate persistent charge 
and spin currents in a mesoscopic ring. The numerical results are 
illustrated in Section III. Finally, in Section IV, we summarize our 
results.

\section{The model and Theoretical formulation}

\subsection{The Model}

Let us start by referring to Fig.~\ref{ring}, where a mesoscopic ring is
subject to an AB flux $\phi$ (measured in unit of the elementary flux
quantum $\phi_0=ch/e$). Within a TB framework the Hamiltonian for such
a $N$-site ring looks in the form,
\begin{eqnarray}
\mbox{\boldmath $H$} &=& \sum_n \mbox{\boldmath $c_n^{\dagger} 
\epsilon_n c_n$} + \sum_n \left(\mbox{\boldmath $c_{n+1}^{\dagger} t$} 
\,e^{i \theta} \mbox {\boldmath $c_n$} + h.c. \right) \nonumber \\
&-& \sum_n \left(\mbox{\boldmath $c_{n+1}^{\dagger} 
($}i \mbox{\boldmath$\sigma_x)~\alpha$} \cos\varphi_{n,n+1} \,e^{i \theta} 
\mbox{\boldmath $c_n$} + h.c. \right) \nonumber \\
& - & \sum_n \left(\mbox{\boldmath $c_{n+1}^{\dagger} 
($}i \mbox{\boldmath$\sigma_y)~\alpha$} \sin\varphi_{n,n+1} \,e^{i \theta} 
\mbox{\boldmath $c_n$} + h.c. \right) \nonumber \\
& + & \sum_n \left(\mbox{\boldmath $c_{n+1}^{\dagger} 
($}i \mbox{\boldmath$\sigma_y)~\beta$} \cos\varphi_{n,n+1} \,e^{i \theta} 
\mbox{\boldmath $c_n$} + h.c. \right) \nonumber \\
& + & \sum_n \left(\mbox{\boldmath $c_{n+1}^{\dagger} 
($}i \mbox{\boldmath$\sigma_x)~\beta$} \sin\varphi_{n,n+1} \,e^{i \theta} 
\mbox{\boldmath $c_n$} + h.c. \right) 
\label{equ1}
\end{eqnarray}
where, $n=1$, $2$, $\dots$, $N$ is the site index along the azimuthal
direction $\varphi$ of the ring. The other factors in Eq.~\ref{equ1}
are as follows.\\
\mbox{\boldmath $c_n^{\dagger}$}=$\left(\begin{array}{cc}
c_{n \uparrow}^{\dagger} & c_{n \downarrow}^{\dagger} 
\end{array}\right);$
\mbox{\boldmath $c_n$}=$\left(\begin{array}{c}
c_{n \uparrow} \\
c_{n \downarrow}\end{array}\right);$
\mbox{\boldmath $\epsilon_n$}=$\left(\begin{array}{cc}
\epsilon_{n\uparrow} & 0 \\
0 & \epsilon_{n\downarrow} \end{array}\right);$ \\
\mbox{\boldmath $t$}=$t\left(\begin{array}{cc}
1 & 0 \\
0 & 1 \end{array}\right);$ 
\mbox{\boldmath $\alpha$}=$\left(\begin{array}{cc}
\alpha & 0 \\
0 & \alpha \end{array}\right);$
\mbox{\boldmath $\beta$}=$\left(\begin{array}{cc}
\beta & 0 \\
0 & \beta \end{array}\right);$ \\
\mbox{\boldmath $\sigma_{x}$}=$\left(\begin{array}{cc}
0 & 1 \\
1 & 0 \end{array}\right);$
\mbox{\boldmath $\sigma_{y}$}=$\left(\begin{array}{cc}
0 & -i \\
i & 0 \end{array}\right);$
\mbox{\boldmath $\sigma_{z}$}=$\left(\begin{array}{cc}
1 & 0 \\
0 & -1 \end{array}\right)$. \\
~\\
\noindent
Here $\epsilon_{n\sigma}$ is the site energy of an electron at the
site $n$ of the ring with spin $\sigma$ ($\uparrow,\downarrow$). $t$ 
is the nearest-neighbor hopping integral and $\theta=2\pi \phi/N$ is the 
phase factor due to AB flux $\phi$ threaded by the ring. $\alpha$ and 
$\beta$ are the isotropic nearest-neighbor transfer integrals which 
measure the strengths of Rashba and Dresselhaus SO couplings, respectively, 
and $\varphi_{n,n+1}=\left(\varphi_n+\varphi_{n+1}\right)/2$, where 
$\varphi_n=2\pi(n-1)/N$. \mbox{\boldmath $\sigma_x$},
\mbox{\boldmath $\sigma_y$} and \mbox{\boldmath $\sigma_z$} are the Pauli 
spin matrices. $c_{n \sigma}^{\dagger}$ ($c_{n \sigma}$) is the creation 
(annihilation) operator of an electron at the site $n$ with spin $\sigma$ 
($\uparrow,\downarrow$).

\subsection{Calculation of persistent charge current}

To get persistent charge current, we begin with the basic equation
of charge current operator \mbox{\boldmath $J_{c}$} in terms of the
velocity operator \mbox{\boldmath ${\dot{x}}$} as, 
\begin{equation}
\mbox{\boldmath $J_{c}$}=\frac{1}{N} e \mbox{\boldmath ${\dot{x}}$}
\label{equ2}
\end{equation}
where, the displacement operator \mbox{\boldmath ${x}$} is expressed in
the form \mbox{\boldmath ${x}$}=$\sum\limits_n \mbox{\boldmath$c_n^{\dagger} 
\,n \,c_n$}$. From Eq.~\ref{equ2} we can write,
\begin{eqnarray}
\mbox{\boldmath $J_{c}$} &=& \frac{e}{N} \frac{1}{i\hbar} 
\left[\mbox{\boldmath ${x}$},\mbox{\boldmath ${H}$}\right] \nonumber \\
 &=& \frac{2\pi i e}{Nh}\left[\mbox{\boldmath ${H}$},\mbox{\boldmath ${x}$}
\right].
\label{equ3}
\end{eqnarray}
Substituting \mbox{\boldmath ${x}$} and \mbox{\boldmath ${H}$} in 
Eq.~\ref{equ3}, and, simplifying it we get final expression of the
charge current operator in the form,
\begin{eqnarray}
\mbox{\boldmath $J_{c}$} &=& \frac{2\pi i e}{N}\sum_n \left(
\mbox{\boldmath $c_n^{\dagger} t_{\varphi}^{\dagger\,n,n+1}$} 
\mbox{\boldmath $c_{n+1}$} \,e^{-i \theta}\right. \nonumber \\
 & - & \left. \mbox{\boldmath $c_{n+1}^{\dagger} t_{\varphi}^{n,n+1}$} 
\mbox{\boldmath $c_n$}\,e^{i \theta} \right) 
\label{equ4}
\end{eqnarray}
where, the matrix elements of \mbox{\boldmath$t_{\varphi}^{n,n+1}$} 
are as follows.
\begin{eqnarray}
\mbox{\boldmath $t_{\varphi}^{n,n+1}$}_{1,1} &=& t \nonumber \\
\mbox{\boldmath $t_{\varphi}^{n,n+1}$}_{1,2} &=& -i\,\alpha \, 
e^{-i\varphi_{n,n+1}}+\beta \, e^{i\varphi_{n,n+1}} \nonumber \\
\mbox{\boldmath $t_{\varphi}^{n,n+1}$}_{2,1} &=& -i\,\alpha \, 
e^{i\varphi_{n,n+1}}-\beta \, e^{-i\varphi_{n,n+1}} \nonumber \\
\mbox{\boldmath $t_{\varphi}^{n,n+1}$}_{2,2} &=& t \nonumber 
\end{eqnarray}
Therefore, for a particular eigenstate $|\psi_k\rangle$ the persistent
charge current becomes,
\begin{equation}
J_c^k=\langle \psi_k|\mbox{\boldmath$J_{c}$}|\psi_k\rangle
\label{equ5}
\end{equation}
where $|\psi_k\rangle=\sum\limits_p a_{p,\uparrow} |p\uparrow\rangle
+ a_{p,\downarrow} |p\downarrow\rangle$. Here $|p\uparrow\rangle$'s 
and $|p\downarrow\rangle$'s are the Wannier states and $a_{p,\uparrow}$'s 
and $a_{p,\downarrow}$'s are the corresponding coefficients. After
simplification of Eq.~\ref{equ5}, the final expression of charge current 
looks like,
\begin{eqnarray}
J_c^k = \frac{2\pi i e}{N} \sum_n \left\{t\, a_{n,\uparrow}^* 
a_{n+1,\uparrow}\, e^{-i\theta} - t\, a_{n+1,\uparrow}^* a_{n,\uparrow} 
\,e^{i\theta}\right\} \nonumber \\
+\frac{2\pi i e}{N} \sum_n \left\{t\, a_{n,\downarrow}^* 
a_{n+1,\downarrow}\,e^{-i\theta} - t\, a_{n+1,\downarrow}^* 
a_{n,\downarrow}\, e^{i\theta}\right\} \nonumber \\
+\frac{2\pi i e}{N} \sum_n \left\{\left(i\alpha e^{-i\varphi_{n,n+1}} 
-\beta e^{i\varphi_{n,n+1}}\right)a_{n,\uparrow}^* a_{n+1,\downarrow}
e^{-i\theta} \right. \nonumber \\
+\left. \left(i\alpha e^{i\varphi_{n,n+1}}+\beta
e^{-i\varphi_{n,n+1}}\right) a_{n+1,\downarrow}^* a_{n,\uparrow} 
e^{i\theta}\right\} \nonumber \\
+\frac{2\pi i e}{N} \sum_n \left\{\left(i\alpha e^{i\varphi_{n,n+1}} 
+\beta e^{-i\varphi_{n,n+1}}\right)a_{n,\downarrow}^* a_{n+1,\uparrow}
e^{-i\theta} \right. \nonumber \\
+\left. \left(i\alpha e^{-i\varphi_{n,n+1}}-\beta 
e^{i\varphi_{n,n+1}}\right) a_{n+1,\uparrow}^* a_{n,\downarrow}
e^{i\theta}\right\}. 
\label{equ6}
\end{eqnarray}
The persistent charge current can also be determined in some other ways
as available in literature. Probably the simplest way of determining 
charge current is the case where first order derivative of ground state 
energy with respect to AB flux $\phi$ is taken into account. Mathematically 
we can write $J_{c}=-\partial E_0(\phi)/\partial \phi$, where $E_0(\phi)$ 
is the total energy for a particular electron filling. But, in our present 
scheme (Eq.~\ref{equ6}), the so-called second quantized approach, there 
are some advantages compared to other available procedures. Firstly, we 
can easily measure charge current in any branch of a complicated network. 
Secondly, the determination of individual responses in separate branches 
helps us to elucidate the actual mechanism of electron transport in a 
more transparent way. 

\subsection{Calculation of persistent spin current}

In order to calculate persistent spin current we start with the following
relation,
\begin{equation}
\mbox{\boldmath $J_{s}$}=\frac{1}{2N}\left(\mbox{\boldmath ${\sigma}$}
\mbox{\boldmath ${\dot{x}}$}+ \mbox{\boldmath ${\dot{x}}$} \mbox{\boldmath 
${\sigma}$}\right)
\label{equ7}
\end{equation}
where, $\mbox{\boldmath ${\sigma}$}=\{\mbox{\boldmath ${\sigma_x}$},
\mbox{\boldmath ${\sigma_y}$},\mbox{\boldmath ${\sigma_z}$}\}$. Therefore,
the polarized spin current operator along the quantized direction ($+Z$)
becomes,
\begin{equation}
\mbox{\boldmath $J_{s}^z$}=\frac{1}{2N}\left(\mbox{\boldmath ${\sigma_z}$}
\mbox{\boldmath ${\dot{x}}$}+ \mbox{\boldmath ${\dot{x}}$} \mbox{\boldmath 
${\sigma_z}$}\right).
\label{equ8}
\end{equation}
Substituting \mbox{\boldmath ${\dot{x}}$} in Eq.~\ref{equ8} and expanding
it the spin current operator gets the form,
\begin{eqnarray}
\mbox{\boldmath $J_{s}^z$} &=& \frac{i\pi}{N}\sum_n \left(
\mbox{\boldmath $c_n^{\dagger}\sigma_z t_{\varphi}^{\dagger\,n,n+1}$} 
\mbox{\boldmath $c_{n+1}$} \,e^{-i \theta}\right. \nonumber \\
& - &\left. \mbox{\boldmath $c_{n+1}^{\dagger}\sigma_z t_{\varphi}^{n,n+1}$} 
\mbox{\boldmath $c_n$}\,e^{i \theta} \right) \nonumber \\
&+& \frac{i\pi}{N}\sum_n \left(\mbox{\boldmath $c_n^{\dagger} 
t_{\varphi}^{\dagger\,n,n+1} \sigma_z$} \mbox{\boldmath $c_{n+1}$} 
\,e^{-i \theta}\right. \nonumber \\
& - &\left. \mbox{\boldmath $c_{n+1}^{\dagger} t_{\varphi}^{n,n+1} \sigma_z$} 
\mbox{\boldmath $c_n$}\,e^{i \theta} \right). 
\label{equ9}
\end{eqnarray}
Using the same prescription, as illustrated in Eq.~\ref{equ5}, we reach 
the final expression of persistent spin current for $k$-th eigenstate as,
\begin{eqnarray}
J_s^{z,k} &=& \frac{2\pi i t}{N}\sum_n\left\{a_{n,\uparrow}^*a_{n+1,\uparrow}
\,e^{-i\theta} - a_{n+1,\uparrow}^* a_{n,\uparrow} \,e^{i\theta}\right\}
\nonumber \\
&-&\frac{2\pi i t}{N} \sum_n \left\{a_{n,\downarrow}^* a_{n+1,\downarrow}
\,e^{-i\theta} - a_{n+1,\downarrow}^* a_{n,\downarrow} \,e^{i\theta}\right\}.
\nonumber \\
\label{equ44}
\end{eqnarray}
In our presentation we refer the polarized spin current $J_s^{z,k}$ as 
$J_s^k$ for the sake of simplicity.

In the present work we examine all the essential features of persistent
charge current, spin current and related issues at absolute zero temperature
and choose the units where $c=h=e=1$. Throughout our numerical work we fix 
$t=1$ and measure the energy scale in unit of $t$.

\section{Numerical results and discussion}

\subsection{Energy spectra}

\begin{figure}[ht] 
{\centering \resizebox*{7.75cm}{12cm}{\includegraphics{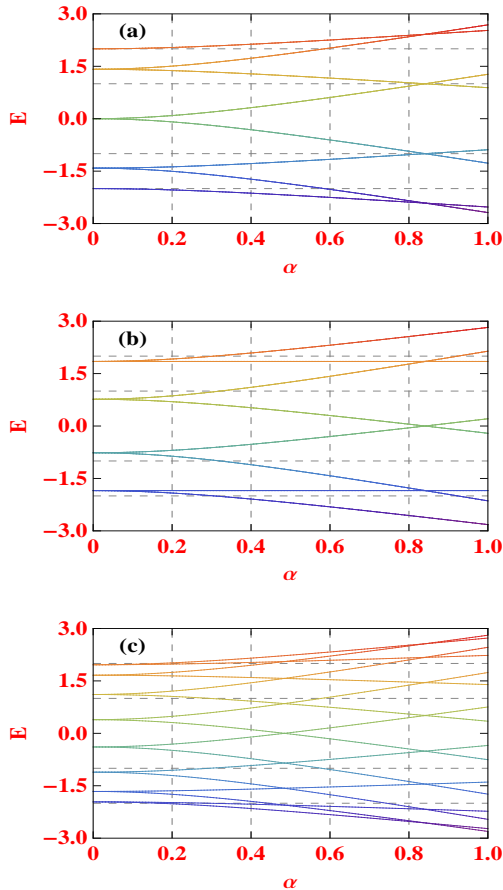}}\par}
\caption{(Color online). $E$-$\alpha$ characteristics for a $8$-site 
ordered ring, where (a), (b) and (c) correspond to $\phi=0$, $0.5$ and 
$0.25$, respectively. For all these cases $\beta$ is fixed at $0$.}
\label{energy}
\end{figure}

To make this present communication a self contained study let us first
start with the energy spectrum of a mesoscopic ring considering SO
interaction for some typical values of AB flux $\phi$ threaded by the
ring. In Fig.~\ref{energy} we present the variation of energy levels
of an ordered $8$-site ring as a function of Rashba SO coupling strength
$\alpha$, where (a), (b) and (c) correspond to $\phi=0$, $\phi_0/2$ and
$\phi_0/4$, respectively. For all these spectra, Dresselhaus SO coupling 
is set at $0$. Now we analyze the behavior of energy levels for the 
three different cases of $\phi$. Case-I: $\phi=0$. When $\alpha=0$, 
the eigenvalues are four-fold degenerate, except the lowest and highest 
eigenvalues those are two-fold degenerate. As the SO interaction is 
switched on ($\alpha \ne 0$) the four-fold degenerate energy levels 
split and provide two-fold Kramers degeneracy. With the increase of 
SO coupling strength, splitting of these energy levels becomes larger 
which is clearly seen from the energy spectrum. The appearance of 
two-fold degenerate energy level/levels at one edge or both edges
of an energy spectrum solely depends on the ring size $N$. If $N$ is
odd, a single two-fold degenerate energy level appears at the top of 
the spectrum. On the other hand, if $N$ is even, in each side of the 
energy spectrum a single two-fold degenerate energy level is obtained 
(Fig.~\ref{energy}(a)).
Case-II: $\phi=\phi_0/2$. At $\alpha=0$, the energy levels are four-fold
degenerate. They get splitted in the presence of SO coupling providing
two-fold Kramers degeneracy. Depending on ring size $N$, here also 
two-fold degenerate energy levels are obtained when SO coupling strength 
is set at zero. When $N$ is odd, a two-fold degenerate energy level 
appears at the bottom of the energy spectrum (opposite to the case of 
$\phi=0$). While, no two-fold degenerate energy levels at $\alpha=0$ 
appears when $N$ becomes an even number (Fig.~\ref{energy}(b)). 
Case-III: $\phi=\phi_0/4$. For any other values of $\phi$, the energy 
levels are two-fold degenerate only when SO coupling is zero, while they 
are non-degenerate when SO coupling is turned on (Fig.~\ref{energy}(c)). 
Exactly similar spectra are obtained when the energy levels for an 
ordered ring are plotted as a function of Dresselhaus SO coupling 
considering $\alpha=0$.  

\subsection{Determination of $\alpha$ and $\beta$ by measuring persistent
charge current}

The existence of dissipationless charge current in a mesoscopic ring 
in presence of magnetic flux is a well-known phenomenon. But at zero
magnetic flux the issue of developing persistent charge 
current~\cite{loss,liu} in a mesoscopic ring solely by SO interaction 
provides a key idea of measuring SO fields. To the best of our knowledge 
this approach of measuring SO coupling has remain unaddressed so far. 

To establish persistent charge current in a mesoscopic ring in the 
absence of traditional AB flux, we consider a ring subject to magnetic
impurities. To get a magnetically disordered ring, we choose 
$\epsilon_{n\uparrow}$ randomly from a ``Box" distribution function of 
width $W$, and, set $\epsilon_{n\downarrow}=-\epsilon_{n\uparrow}$ for 
all $n$. It reveals that the localized magnetic moments, placed at 
different atomic sites of the ring, are aligned along the quantized 
($+Z$) direction. In such a ring, SO interaction can produce persistent 
charge current even in the absence of AB flux $\phi$.

At absolute zero temperature ($T=0$k), net persistent charge current for 
a ring described with $N_e$ electrons can be determined by taking the
\begin{figure}[ht] 
{\centering \resizebox*{8cm}{12cm}
{\includegraphics{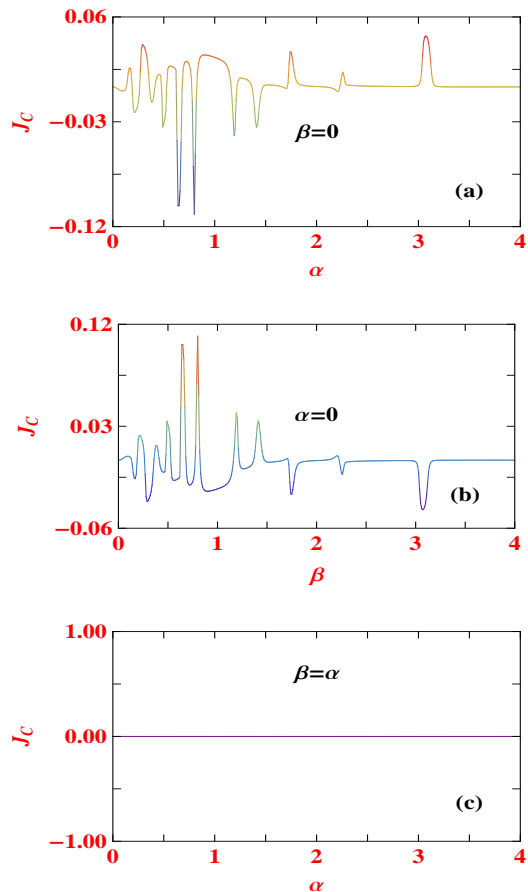}}\par}
\caption{(Color online). Persistent charge current ($J_c$) of a 
$60$-site magnetically disordered ring in the quarter-filled case
for different values of $\alpha$ and $\beta$ when AB flux $\phi=0$.}
\label{chargecurrent}
\end{figure}
sum of individual contributions from the lowest $N_e$ energy eigenstates.
Hence, for $N_e$ electron system total charge current becomes,
\begin{equation}
J_c=\sum_k^{N_e} J_c^k.
\label{pcc}
\end{equation}
In Fig.~\ref{chargecurrent} we establish the variation of persistent
charge current $J_c$ of a magnetically disordered ($W=1$) $60$-site ring
in the quarter-filled case ($N_e=30$) for different values of $\alpha$
and $\beta$ when conventional electromagnetic flux through the ring is
set at zero. From the spectra we see that depending on the values of 
$\alpha$ and $\beta$, charge current show several complex behavior
and we justify them in the following ways. Case-I: $\beta=0$.
At $\alpha=0$, no charge current appears in the ring. But as long as
Rashba SO coupling is turned on charge current is established in the
ring (Fig.~\ref{chargecurrent}(a)). When an electron circulates in a 
\begin{figure}[ht] 
{\centering \resizebox*{8cm}{12cm}
{\includegraphics{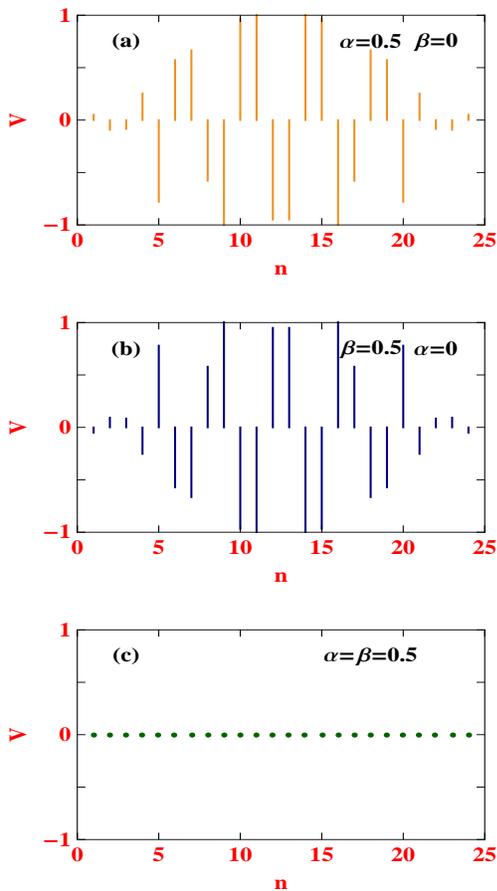}}\par}
\caption{(Color online). Velocity ($V$) of an electron in different energy
eigenstates ($n$) of a $12$-site magnetically disordered ring for some 
typical values of $\alpha$ and $\beta$ when AB flux $\phi=0$.}
\label{velocity}
\end{figure}
magnetically disordered ring in presence of SO coupling, it acquires
a geometric phase the so-called Berry phase~\cite{berry}. This 
geometrical phase provides the dissipationless charge current in the 
ring. Accordingly, in the absence of SO coupling charge current does 
not appear. The existence of dissipationless charge current in presence
of SO coupling can also be verified in other way by studying velocity
distribution of different energy eigenstates. As illustrative example,
in Fig.~\ref{velocity}(a) we plot the velocity ($V$) of an electron in 
different energy eigenstates ($n$) of a $12$-site magnetically disordered
($W=1$) ring considering $\alpha=0.5$ and $\beta=0$. To reveal the $V$-$n$ 
spectrum more transparently we choose such a small sized ring ($N=12$). 
The spectrum shows that the velocity of an electron changes significantly 
as we go on from one eigenstate to other, and also, for some energy levels 
electrons are moving in one direction and for other levels electrons are 
rotating in the opposite direction. Since the net charge current is 
obtained by taking the sum of individual contributions from the lowest 
$N_e$ energy levels, a finite non-zero charge current appears in presence 
of $\alpha$. Therefore, it can be manifested that Rashba SO interaction 
can induce a dissipationless charge current in a magnetic disordered ring 
even in the absence of traditional AB flux $\phi$. Case-II: $\alpha=0$. 
Now we consider the effect of Dresselhaus SO
\begin{figure}[ht] 
{\centering \resizebox*{8cm}{12cm}{\includegraphics{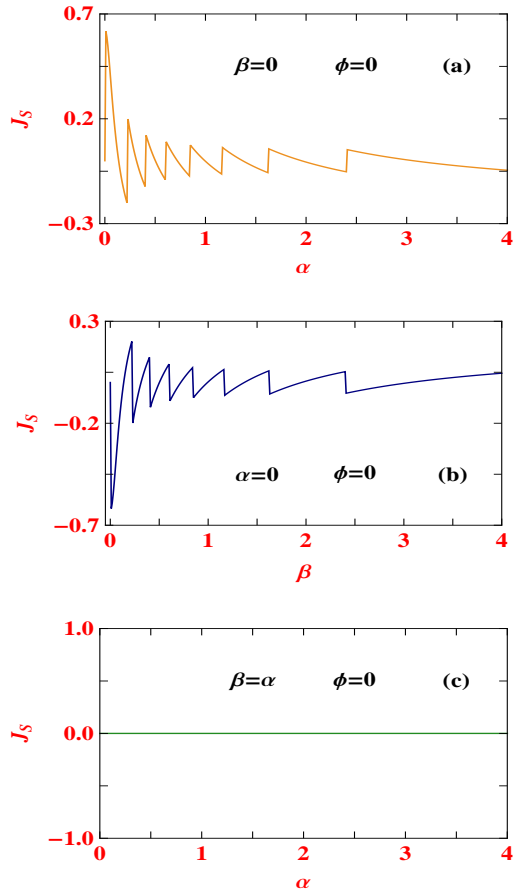}}\par}
\caption{(Color online). Persistent spin current ($J_s$) of a $40$-site 
ordered half-filled ring for different values of $\alpha$ and $\beta$ when
AB flux $\phi=0$.}
\label{spincurrent1}
\end{figure}
interaction in a magnetic disordered ring when other SO field is set at
zero. The nature of charge current is
presented in Fig.~\ref{chargecurrent}(b). Similar to the above case here
also the charge current disappears when $\beta=0$. While, in the presence 
of Dresselhaus SO coupling non-vanishing charge current appears and it 
shows exactly opposite behavior of the $J_c$-$\alpha$ characteristic 
curve (Fig.~\ref{chargecurrent}(a)). The non-vanishing behavior of charge
current in presence of $\beta$ can be easily justified from the $V$-$n$
spectrum plotted in Fig.~\ref{velocity}(b). Comparing the $V$-$n$ spectra
given in Figs.~\ref{velocity}(a) and (b), opposite nature of charge current
in the cases of $\alpha$ and $\beta$ is clearly understood. From these 
spectra it is observed that in presence of $\alpha$ the velocity of an 
electron in a particular energy eigenstate is exactly identical in 
magnitude and opposite in sign with the velocity of an electron in that 
particular eigenstate when $\alpha$ is replaced by $\beta$. It provides
opposite currents in the two different cases of SO fields. Case-III: 
$\alpha=\beta$. The situation becomes very much interesting when both
SO coupling strengths are identical in magnitude. In this particular case 
charge current completely disappears. The result is shown in 
Fig.~\ref{chargecurrent}(c) and the vanishing behavior of charge current 
is clearly understood from the $V$-$n$ spectrum given in 
Fig.~\ref{velocity}(c). For the typical case when $\alpha=\beta$, velocity 
of an electron drops exactly to zero for all the energy eigenstates which 
provides vanishing charge current. This phenomenon leads to an important 
idea for the determination of SO fields. It is well known that in a 
material Rashba strength can be tuned by applying electric fields from gate 
or by monitoring the density of electrons. Hence, for a particular sample 
subject to Rashba and Dresselhaus SO interactions, vanishing charge current 
can be obtained by properly tuning the Rashba SO coupling making its 
strength identical to the Dresselhaus SO coupling. This, on the other hand, 
determines the Dresselhaus SO coupling. 

In presence of finite AB flux through the ring this approach cannot be 
used to determine SO fields, since then charge current will not vanish 
for the particular case when Rashba and Dresselhaus SO coupling strengths 
are identical to each other. Accordingly, some other methods have to be 
utilized for the determination of these fields. In a recent work~\cite{san2} 
we have shown that by estimating conductance minimum, calculated in terms of 
Drude weight, a closely related parameter that characterizes conducting
nature of a system as originally noted by Kohn~\cite{kohn}, SO strengths 
can be determined.

\subsection{Determination of $\alpha$ and $\beta$ by measuring persistent
spin current}

In this sub-section we establish another approach of estimating Rashba and
Dresselhaus SO fields by measuring persistent spin current in a mesoscopic
ring instead of charge current. In order to understand the meaning of a 
spin current, assume a current passes through a channel which contains 
only up-spin polarized electrons. Now include a similar current with it 
which flows in the opposite direction and contains only down-spin 
polarized electrons. As a result, the net transfer of electrons across 
any cross section of the channel becomes zero, but it leads to a current 
of spins which is the so-called spin current. It differs from a charge 
current in two aspects. First, it is associated with a flow of angular 
momentum. Second, it maintains the time-reversal symmetry~\cite{sharma}.

Here we show that a non-magnetic mesoscopic ring with a SO interaction 
can provide a dissipationless pure spin current even in the absence of 
conventional electromagnetic flux through the ring and it provides an
idea of measuring SO coupling strengths.

At absolute zero temperature ($T=0$k), net persistent spin current in 
a mesoscopic ring for a particular filling can be obtained by taking
the sum of individual contributions from the energy levels with energies
less than or equal to Fermi energy $E_F$. Therefore, for $N_e$ electron
system total spin current becomes,
\begin{equation}
J_s=\sum_k^{N_e} J_s^k.
\label{psc}
\end{equation}
In Fig.~\ref{spincurrent1} we show the variation of persistent spin 
current $J_s$ of an ordered $40$-site ring in the half-field case
($N_e=40$) for different values of $\alpha$ and $\beta$ when AB 
\begin{figure}[ht] 
{\centering \resizebox*{8cm}{12cm}{\includegraphics{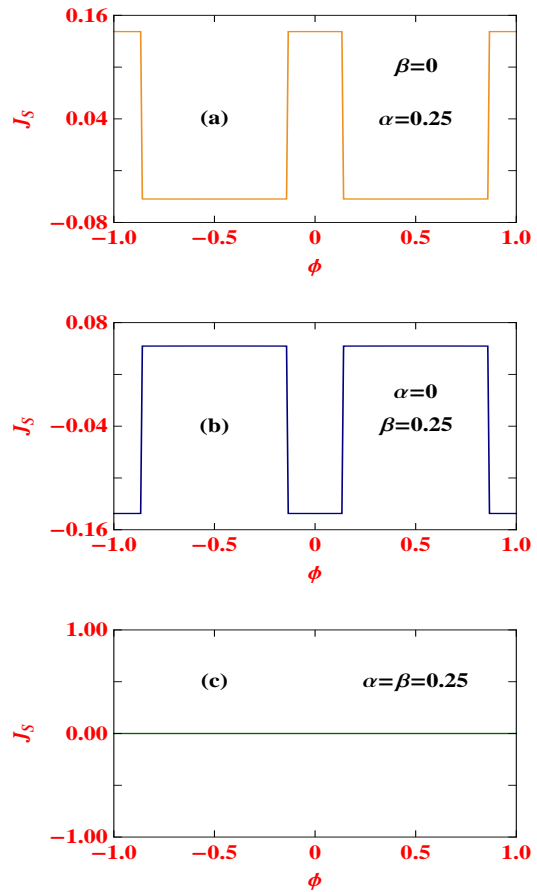}}\par}
\caption{(Color online). Persistent spin current ($J_s$) as a function 
of $\phi$ of a $40$-site ordered half-filled ring for some typical
values of $\alpha$ and $\beta$.}
\label{spincurrent2}
\end{figure}
flux $\phi$ is fixed at zero. Several interesting features are observed
those are implemented as follows.
Case-I: $\beta=0$. At $\alpha=0$, the ring does not support any spin
current. While, a non-vanishing spin current appears as long as Rashba
SO interaction is turned on (Fig.~\ref{spincurrent1}(a)). Like a driving 
force for the case of persistent charge current, one also looks for the 
analogous driving force in the case of pure persistent spin current. It 
is the SO interaction which plays the role of spin driving force and leads
to a dissipationless pure spin current. This can be justified in the 
following way. Let us consider an electron with spin $\sigma$ ($\uparrow,
\downarrow$) circulating in the ring subject to a SO interaction only.
In presence of SO interaction, the spin of this electron precesses and 
gets a geometric phase when the electron comes back to its initial 
position. This geometric phase is the so-called spin Berry 
phase~\cite{gel1,gel2}, and, for an electron with spin $\sigma$ traversing
along the ring it can be expressed as: $\chi_{\sigma}=\sigma\pi$, where 
$\sigma=\pm$ for $\sigma=\uparrow,\downarrow$. The spin Berry phase 
$\chi_+$ provides a clockwise polarized persistent spin current, while
the phase $\chi_-$ induces an anti-clockwise spin current with the 
polarization exactly opposite to the earlier one since the time-reversal
symmetry is preserved for the ring~\cite{loss}. It reveals a pure
persistent spin current. Case-II: $\alpha=0$. When $\beta=0$, no spin
current appears in the ring since in this case there is no driving force 
for generating the current. On the other hand, a dissipationless spin
current is appeared when $\beta$ is finite (Fig.~\ref{spincurrent1}(b)). 
From the spectrum (Fig.~\ref{spincurrent1}(b)) we see that the nature
of spin current is exactly opposite in nature compared to the 
$J_s$-$\alpha$ characteristic curve (Fig.~\ref{spincurrent1}(a)). This
feature can be implemented exactly in the similar way as studied in
the previous section. Thus, Rashba or Dresselhaus SO field can induce
a pure spin current even in the absence of an external magnetic field 
or a magnetic flux. Case-III: $\alpha=\beta$. Finally, when both the two 
spin orbit strengths are identical, spin current drops exactly 
to zero. It is given in Fig.~\ref{spincurrent1}(c). The vanishing nature 
of spin current in this particular case can be clearly understood since 
Rashba and Dresselhaus SO interactions induce spin currents exactly 
identical in magnitude but their directions are opposite to each other. 
Thus, for a particular material subject to Rashba and Dresselhaus SO 
fields, vanishing spin current is achieved by adjusting the Rashba 
coupling to the Dresselhaus strength. This behavior helps us to predict 
the strengths of these SO fields. 

To make an end of our discussion in a compact form, finally, we 
concentrate on the variation of persistent spin current as a function
of AB flux $\phi$ for some typical values of Rashba and Dresselhaus
SO fields. The results for a $40$-site ordered half-filled ($N_e=40$) 
ring are shown in Fig.~\ref{spincurrent2}. When anyone of the two SO
interactions is finite and other is zero, persistent spin current varies
periodically with $\phi$ exhibiting $\phi_0$ flux-quantum periodicity. 
From these spectra (Figs.~\ref{spincurrent2}(a) and (b)) we also see 
that Rashba and Dresselhaus SO fields lead to spin currents exactly
identical in magnitude but they are polarized in the opposite directions. 
This gives the vanishing behavior of persistent spin current when the 
strengths of these two SO fields are equal (Fig.~\ref{spincurrent2}(c)). 
Interestingly, we see that the vanishing nature of spin current is 
observed even for the non-zero value of AB flux $\phi$.

\section{Closing remarks}

To summarize, we have explored two different possibilities of measuring
the absolute values of Rashba and Dresselhaus spin-orbit fields in a 
single sample. In the first approach we have estimated the strength of 
the SO fields by calculating persistent charge current in a mesoscopic 
ring subject to magnetic impurities. In such a ring SO interaction 
induces persistent charge current even in the absence of conventional
electromagnetic flux through the ring. The charge current completely
disappears when the strengths of both these two SO fields are identical,
and this phenomenon helps us to estimate the SO coupling strengths. 
In the other approach, we have determined the strengths of SO fields
by calculating persistent spin current in a non-magnetic mesoscopic 
ring. We have shown that, even in the absence of conventional AB flux,
SO interaction leads to a persistent spin current. Here also, spin 
current vanishes when Rashba and Dresselhaus SO fields are identical
in magnitude. This on the other hand, gives the possibility of estimating
SO strengths. We hope our numerical results can be observed experimentally.

\vskip 0.3in
\noindent
{\bf\small ACKNOWLEDGMENTS}
\vskip 0.2in
\noindent
It is a pleasure to thank S. Sil, A. Chakrabarti, S. N. Karmakar and 
M. Dey for several helpful conversations.

\end{document}